\documentclass[11pt]{article}
\usepackage{moriond,epsfig}

\def\as{{\alpha}_s}
\def\bas{\bar{\alpha}_s}
\def\MSbar{\relax\ifmmode\overline{\rm MS}\else{$\overline{\rm MS}${ }}\fi}
\def\gam{\gamma}
\def\ee{e^+e^-}

\def\sig{\sigma}
\def\cP{{\cal P}}

\begin{document}
\vspace*{4cm}
\title{RELATING SMALL FEYNMAN AND BJOKEN $x$}

\author{GIUSEPPE MARCHESINI}

\address{University of Milan-Bicocca and INFN, Sezione di
Milano-Bicocca, Milan, It}

\maketitle\abstracts{ This is a working progress report on the attempt
  by Yuri Dokshitzer, Gavin Salam and myself to relate the small-$x$
  behaviour of the anomalous dimensions in the time- and space-like
  cases.  This relation is based on a {\it reciprocity respecting
    equation} we propose}

\section{Motivations}
It is challenging to try to extract physics features from the $10^2$-page
long formula of the three loops space-like anomalous
dimensions~\cite{MVV}. 
That these coefficients could reveal general features of QCD can be
illustrated by the example of coherence of QCD radiation.
Consider the formula for the small-$N$ limit of the time-like
anomalous dimension which, in double logarithmic approximation (DLA), is
given by ($N$ is the Mellin moment conjugate to $x$, the limit $x\!\to\!0$
corresponds to $N\!\to\!0$)  
\begin{equation}
  \label{eq:mult}
\gam^{\rm DLA}_+(N,\as)=\frac{1}{4}\left( \sqrt{N^2+8\bas}-N\right)=
\frac{\bas}{N}-\frac{2\,\bas^2}{N^3}+\cdots\qquad 
\bas=\frac{C_A\,\as}{\pi}\,.
\end{equation}
This formula was derived \cite{soft} by studying multi-soft gluon
distributions and the evolution of jets. Here one discovered QCD
coherence, i.e. cancellation in part of phase space due to destructive
interference leading to angular ordering.  Actually, this important
feature of QCD radiation is implicit in the formula of the two loops
time-like anomalous dimension \cite{CFP}.
Indeed, if in the jet evolution one would take into account the full
kinematically available phase space, instead of the correct two-loop
coefficient $-2/N^3$ in (\ref{eq:mult}) one would obtain $-1/N^3$ and
this is the signal of the need of cancellations in part of phase
space.

The attempt to recognise physics features from the high order
coefficients of the space- and time-like anomalous dimensions is a
long standing Yuri's project that, after the publication of the three
loop space-like anomalous dimension, we revisited. We
studied~\cite{DMS} the large-$x$ region and analysed to what extent
classical soft emission~\cite{LBK} and the reciprocity
relation~\cite{DLY} can explain the structure of the known expansion
coefficients. In this talk I report some further considerations in
this project and I focus on the small-$x$ region.

\section{Reciprocity respecting equation}
The fact that the Bjorken and Feynman variables in DIS and $\ee$
inclusive fragmentation 
\begin{equation}
\label{eq:xes}
  x_B \>=\> \frac{-q^2}{2(Pq)}, \qquad  x_F \>=\> \frac{2(Pq)}{q^2},
\end{equation}
are indicated by the same letter \footnote{ In DIS $q$ is the large
  space-like momentum transferred from the incident lepton to the
  target nucleon $P$.  In $\ee$ annihilation $q$ is the time-like
  total incoming momentum and $P$ the final observed hadron.}  is
certainly not accidental.  These variables are mutually reciprocal:
after the crossing operation $P\to -P$ one $x$ becomes the inverse of
the other (although in both channels $0\le x \le 1$ thus requiring the
analytical continuation).

Such a reciprocity property can be extended to the Feynman diagrams
for the two processes and, in particular, to the contributions from
mass-singularities. Consider, for DIS (S-case) and $\ee$ annihilation
(T-case), the skeleton structure~\cite{KKST} of Feynman graphs in
axial gauge and the mass-singularities phase space ordering:

\vspace{0.7 cm}
\begin{minipage}{0.40\textwidth}
    \epsfig{file=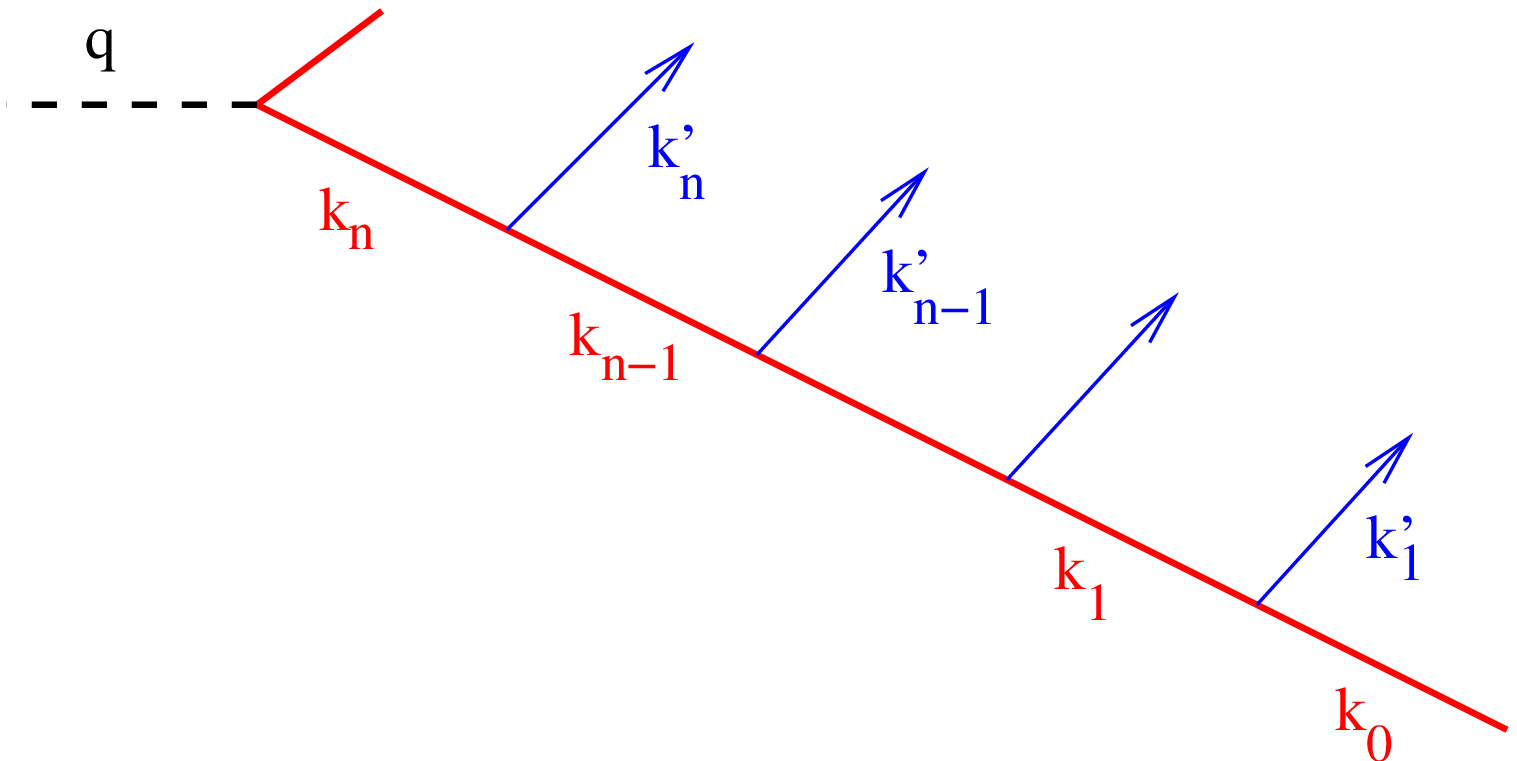,width=0.9\textwidth}
\end{minipage}
\begin{minipage}{0.55\textwidth}
\begin{eqnarray}
\label{eq:orders}
&&\mbox{S-case:}\quad -k^2_{i\!-\!1} <-k^2_{i}\>\frac{k_{i\!-\!1+}}{k_{i+}}
=-k^2_{i}\> {z_i^{-1}}\\
\label{eq:ordert}
&&\mbox{T-case:}\quad  k^2_{i\!-\!1} < k^2_{i}\>\frac{k_{i\!-\!1+}}{k_{i+}}
=k^2_{i}\> { z_i}
\end{eqnarray}
\end{minipage}

\vspace{0.7 cm}
\noindent
The same Feynman graphs are contributing and, going from S- to
T-channel, the mass singularities are obtained by reciprocity: change
$z$ into $1/z$ and the momentum $k$ from space-like to time-like.

This fact is at the origin of the Drell-Levy-Yan relation \cite{DLY}
which has been largely used in order to obtain the time-like anomalous
dimensions from the space-like ones \cite{CFP,SV}.
When dimensional regularization is used, this simple kinematical
reciprocity relation is corrupted: the coefficient functions and
parton distributions in the S- and T-channel differ by factors
$z^{-2\epsilon}$ (from the phase space) and $(1\!-\!\epsilon)$ from
spin averages. However these corrections do not lead to really new
structures but are mostly related to the anomalous dimensions to lower
order.

Could this $\epsilon$-corruption be a peculiar artifact of the
calculation in dimensional regularization so that, at the end of the
calculation, reciprocity is restored?  This question was also raised 
by Stratmann and Vogelsang~\cite{SV}. 
In the following we explore this. More precisely we assume that parton
distributions in the two channels simply resum mass-singularities from
the ordered phase space (\ref{eq:orders},\ref{eq:ordert}) and we
neglect regularization subtleties.

We introduce the probability $D_\sig(N,\kappa^2)$ to find a parton
with virtuality $\sig k^2$ up to $\kappa^2$ with $\sig\!=\!-1$ for
the S-case and $\sig\!=\!1$ for the T-case.  The ordering
(\ref{eq:orders},\ref{eq:ordert}) gives rise to the following {\it
reciprocity respecting equation}
\begin{equation}
\label{eq:rre}
\kappa^2\partial_{\kappa^2}D_\sig(N,\kappa^2)= 
\gam_\sig(N)\, D_\sig(N,\kappa^2) =
\int_0^1\frac{dz}{z}\,z^N\,P(z,\as)\,D_\sig(N,\kappa^2\,z^\sig)\,,
\quad\sig=\pm1\,.
\end{equation}
The difference between the two channels is simply in the fact that the
virtuality of the integrated parton distribution is
$\kappa^2\,z^\sig$, see (\ref{eq:orders},\ref{eq:ordert}). The
splitting function $P(z,\as)$ does not depend on the S- or T-channel
(its Mellin moments are not the anomalous dimensions).  The running
coupling in the splitting function depends \cite{DMS} on the
virtuality in a reciprocity respecting form.
This equation (in general a matrix equation) is non-local: the S-case
with $\sig\!=\!-1$ (the T-case with $\sig\!=\!1$) involves the parton
distribution with virtualities larger (smaller) than $\kappa$.  So it
is not suitable for explicit calculations of the anomalous dimensions,
but to relate them. 

Here I discuss some of the consequences of (\ref{eq:rre}). I neglect
the running coupling dependence so that in $\gam_\sig(N)$ all
beta-function dependent contributions are missed~\footnote{Other
  beta-function dependent contributions to the anomalous dimensions
  are generated by changing coefficient functions.}.  In this case
(\ref{eq:rre}) can be solved as (see also~\cite{MM})
\begin{equation}
  \label{eq:cP}
\gam_\sig(N)=\cP(N+\sig\gam_\sig(N))\,,\qquad 
\cP(n)=\int_0^1\frac{dz}{z}\,z^n\, P(z,\as)\,.  
\end{equation}
The case of large-$x$ (corresponding to large-$N$) has been discussed
in \cite{DMS}. Here the Mellin transformed $x$-dependent anomalous
dimensions $\tilde\gam_\sig(x)$ have the expansion
\begin{equation}
  \label{eq:largex}
\tilde\gam_\sig(x)=
A_\sig\frac{x}{(1-x)_+}+B_\sig\delta(1-x)+C_\sig\ln(1-x)+D_\sig+\cdots
\end{equation}
These four coefficients are non-vanishing only for the diagonal
matrix-elements and are given by expansions in $\as$. By using the
reciprocity respecting equation (\ref{eq:cP}) one obtains, 
for the quark and gluon matrix elements
\begin{equation}
  \label{eq:largex1}
  A_\pm=A\,,\qquad
  B_\pm=B\,,\qquad
  C_\sig=-\sig A^2\,,\qquad
  D_\sig=-\sig AB\,.
\end{equation}
For both the quark and gluon channel, these relations are satisfied
(neglecting the beta-function contribution) in the S- and T-case at
2-loop level~\cite{CFP} and in the S-case~\cite{MVV} at 3-loop.

Recently it has actually been shown~\cite{MiMV} that (\ref{eq:rre})
holds for the whole of the $\sigma=1$ 3-loop non-singlet splitting
function.

\section{Smal- $x$ case}
For $N\to0$ we consider a single anomalous dimension for the S- and
T-channel, essentially the gluon-gluon case.  
The most singular terms of $\gam_-(N)$ are given by the BFKL
formula~\cite{BFKL}, an expansion in $(\bas/N)^p$.  Next-to-BFKL
contributions $\bas\,(\bas/N)^{p}$ are also known \cite{N-BFKL}.
Therefore, in the S-case, all singular terms $\bas^p/N^{k}$ with $k>p$
fully cancel. To recall the physics behind this result, consider the
leading order term.

\paragraph{DLA.} 
There is a single DLA term in the space-like anomalous dimension
\begin{equation}
\label{eq:DLAs}
\gam_-^{\rm DLA}(N)=\frac{\bas}{N}.
\end{equation}
The fact that all other singular terms vanish, results from
cancellations (coherence) in the mass-singularity phase space
(\ref{eq:orders}) leaving, to this order, transverse momentum
ordering.

Given $\gam_-^{\rm DLA}(N)$ one derives, from the reciprocity equation
(\ref{eq:rre}), the corresponding time-like anomalous dimension
$\gam_+^{\rm DLA}(N)$ given in (\ref{eq:mult}) which, as mentioned,
results from angular ordering.  Therefore coherence in the S-case
($k_t$-ordering) implies coherence in the T-case (ordering in the
angle $k_t/k_+$) and this is just the kinematical reciprocity
transformation.

\paragraph{MLLA.} The small-$x$ gluon-gluon space-like anomalous
dimension at one-loop order is
\begin{equation}
\label{eq:Ssmallx}
 \gamma_-(N) \>=\> \frac{\bas}{N} - a\,\bas\,,\qquad 
a=\frac{11}{12}+\frac{n_f}{6N_c^3}\,.
\end{equation}
%
Given (\ref{eq:Ssmallx}), the reciprocity respecting equation
(\ref{eq:rre}) gives the time-like anomalous dimension 
\begin{equation}
\label{eq:smallxSOL}
\gamma_+(N)\>=\>\frac{1}{4}\left(-(N+2a\,\bas)+
\sqrt{(N-2a\,\bas)^2+8\,\bas}\right).
\end{equation}
The first subleading correction $a\,\bas$ to the DLA expression
(\ref{eq:mult}) corresponds to the MLLA result (modulo the missing
term proportional to $\beta_{0}$) obtained by using the fact that {\em
exact}\/ angular ordering does not acquire~\cite{DT} specific soft
corrections from 2-gluon configurations.

\paragraph{Higher order terms.} 
The expansion for the S- and T-case can be organized as follows
according to increasing singularities for $N\to0$
\begin{equation}
  \label{eq:exp}
\gam_-(N)=\sum_{p=1}^\infty\sum_{k=0}^ps_{p\,k}\>\frac{\bas^{p}}{N^k}\,,
\qquad
\gam_+(N)=\sum_{p=1}^\infty\sum_{k=0}^{2p-1}t_{p\,k}\>\frac{\bas^{p}}{N^k}\,,
\end{equation}
The coefficients $s_{p\,k}$ and $t_{p\,k}$ are then arranged into
lines as depicted in the following chart.

\begin{minipage}{0.45\textwidth}
 \epsfig{file=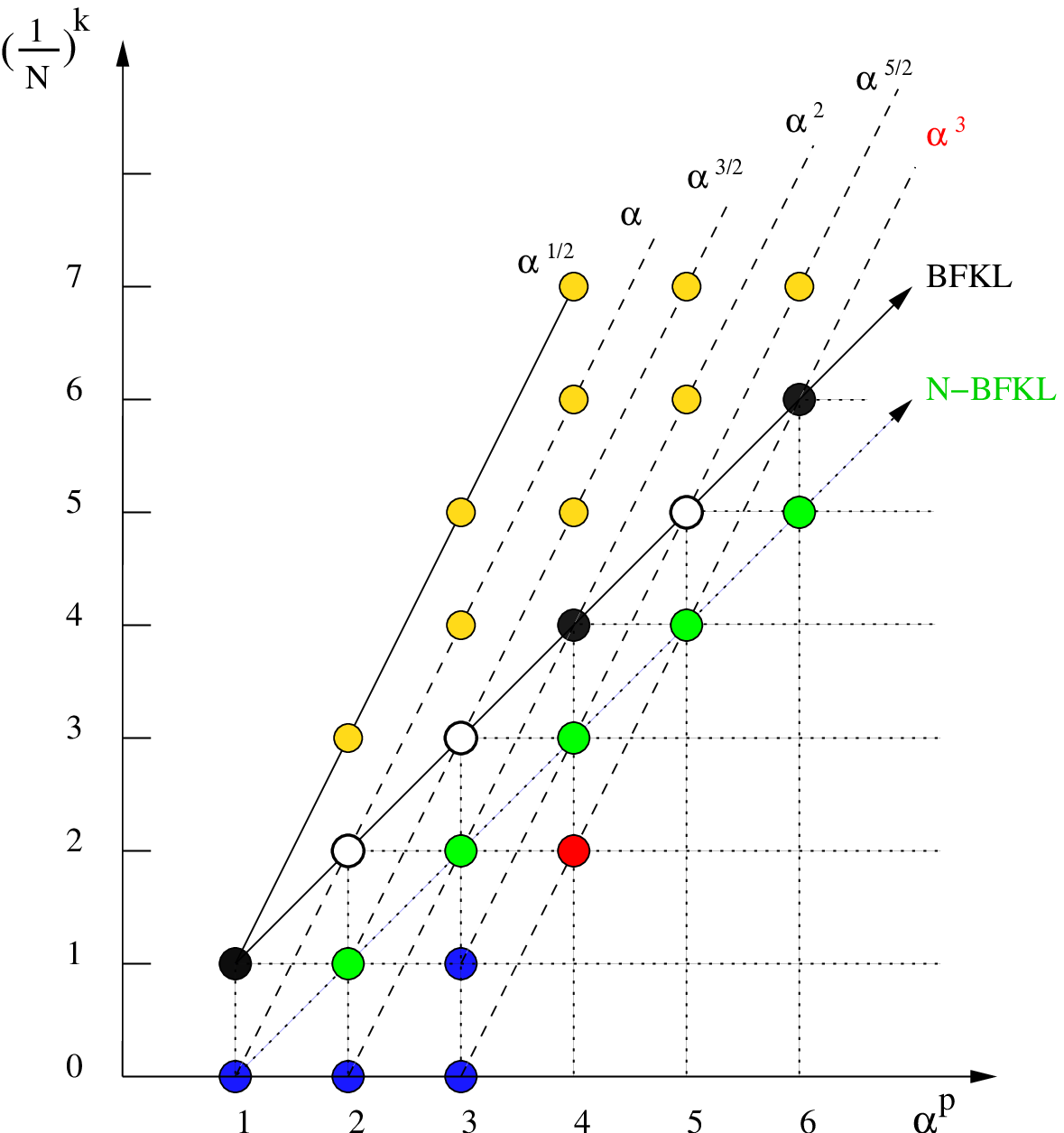,width=\textwidth}
\end{minipage}
\hfill
\begin{minipage}{0.5\textwidth}
  Solid line with black circles shows the BFKL terms ($\bas^p/N^p$).
  Here empty circles stand for BFKL terms that are ``accidentally''
  zero.

\vspace{0.3 cm}

  Line with green circles shows the next-to-BFKL terms
  ($\bas^{p}/N^{p-1}$). 

\vspace{0.3 cm}

Light blue circles mark the other terms known from exact calculations
up to three loops.

\vspace{0.3 cm}

Dashed lines with yellow circles mark series in $\gam_+$ that are
generated by terms of $\gam_-$ on the same lines.  Taking
$N\sim\sqrt{\bas}$ all terms on the same line are of order
$(\sqrt{\bas})^{\tau}$ with $\tau=1,2,\dots$ moving to right.
\end{minipage}

\vspace{0.5cm}
\noindent
The first two series for $\gam_+(N)$ with $\tau=1,2$ are generated by
the $\bas/N$ and $\bas$ terms in $\gamma_-(N)$, see
(\ref{eq:DLAs},\ref{eq:smallxSOL}). Notice that this results from the
fact that the BFKL term of order $\bas^2/N^2$ is ``accidentally''
zero.

\paragraph{Predictions.} 
The known terms of $\gam_-(N)$ allow us to obtain the series of
$\gam_+(N)$ corresponding to the lines $(\sqrt{\bas})^{\tau}$ with
$\tau$ up to $\tau=5$ as follows:
\begin{itemize}
\item the series with $\tau=3$ is generated accounting only for
  next-to-BFKL term $c\,\bas^2/N$ (the $\bas^3/N^3$ BFKL-coefficient
  vanishes).  The corresponding expression for $\gam_+(N)$ is again
  given by (\ref{eq:smallxSOL}) with $8\,\bas$ in the square bracket
  replaced by $8(\bas+ c\,\bas^2)$ and this is just a redefinition of
  the coupling;
\item from the previous cases with $\tau=2,3$, we conclude that the
  fact that both $\bas^2/N^2$ and $\bas^3/N^3$ BFKL-coefficient are
  ``accidentally'' zero (first two white circles in the chart)
  translates into {\em exact}\/ angular ordering \cite{MW,GM}
  according to which no specific soft corrections from 2- and 3-soft
  gluon configurations emerge;

\item the series with $\tau=4$ can be computed from terms in
  $\gam_-(N)$ of order $\bas^2$ (from exact two loop
  results~\cite{CFP}), $\bas^4/N^4$ and $\as^3/N^2$ from BFKL and
  next-to-BFKL respectively;
\item the series with $\tau=5$ can be obtained including $\bas^3/N$
  (from exact three loop results~\cite{MVV}) and $\bas^4/N^3$ from
  next-to-BFKL (BFKL term $\bas^5/N^5$ is accidentally zero);
\item the first series in $\gam_+(N)$ which cannot be computed from
  known space-like results corresponds to $\tau=6$ since the
  coefficient $\bas^4/N^2$ of $\gam_-(N)$ is not known (red circle in
  the chart).
\end{itemize}

\section{Final considerations}

This is a working progress report on the attempt by Yuri Dokshitzer,
Gavin Salam and myself to relate the small-$x$ behaviour of anomalous
dimensions in the time- and space-like cases. I reported here only the
case in which the scale of the running coupling is neglect (no
beta-function contributions~
\footnote{This limit might be most easily be seen in the N=4 SUSY
  limit.}  in $\gam_\pm(N)$).  
The basis is the reciprocity respecting equation (\ref{eq:rre}) which
is deduced by taking into account simply the reciprocity relation in
the mass-singularity phase space (\ref{eq:orders},\ref{eq:ordert}) for
the S- and T-case.  There may be subtleties coming from regularization
and factorization prescriptions used in NLO calculations which go
beyond the analysis of mass-singularities phase space. This is what
happens in the $\MSbar$ calculation, such as extra factors
$z^{-2\epsilon}$ or $(1-\epsilon)$.  On the other hand, these factors
are absent if one uses the Wilson-Polchinski regularization \cite{WP}
scheme (one works in four dimension and introduces a momentum cutoff
and counter-terms to ensure~\cite{BDM} gauge symmetry).
How to rescue reciprocity in terms of renormalization scheme
transformation has been discussed also in~\cite{SV}. An example could
be the scheme used in~\cite{KKST} where, in dimensional
regularization, the $\epsilon\to0$ limit is taken before vituality
integration and then the role of mass-singularity ordering is more
clear.

Our study should allow us to check whether regularization subtleties
could at the end leave uncorrupted the reciprocity relation at the
level of relating physical observables.



\end{document}